# Exoplanets Around Red Giants: Distribution and Habitability

Ruixuan E. Chen[1], Jonathan H. Jiang[2,*], Philip E. Rosen[3], Kristen A. Fahy[2], Yanbei Chen[4]

[1]Arcadia High School, Arcadia, CA 91006
[2]Jet Propulsion Laboratory, California Institute of Technology, Pasadena, CA 91099
[3]Independent Researcher, Vancouver, WA 98662
[4]Burke Institute for Theoretical Physics, California Institute of Technology, Pasadena, CA 91125

*Correspondence: Jonathan.H.Jiang@jpl.nasa.gov

**Abstract**

As the search for exoplanets continues, more are being discovered orbiting Red Giant stars. We use current data from the NASA Exoplanet Archive to investigate planet distribution around Red Giant stars and their presence in the host's habitable zone. As well, we update the power law relation between planet mass and stellar radius found in previous studies and provide more detailed investigations on this topic. Ten Red Giant-hosted exoplanets are found to be in the optimistically calculated habitable zone, five of which are in a more conservatively calculated habitable zone. We believe additional exoplanets can be found in habitable zones around Red Giants using the direct imaging and other methods, along with more powerful detection instrumentation.

## 1. Introduction

In the distant future when our Sun becomes a Red Giant, the habitable zone (HZ) in the Solar System may move towards the outer planets where the moons of Jupiter and Saturn might be candidates for our future generations to live (Sparrman 2022). Near-term considerations also prompt interest in exoplanet and exomoon systems of Red Giant hosts as some of these worlds may presently be in the HZ of their parent star. In this paper we examine data from the NASA Exoplanet Archive, focusing on exoplanets around Red Giant (or Red Subgiant) stars.

When a star leaves the Main Sequence and begins its evolution into the Red Giant Branch (RGB), it undergoes a series of changes. As the fusion of hydrogen progresses in the core of a Main Sequence star, its effective temperature and luminosity increase slowly over time. At the end of a star's Main Sequence stage its core is composed of helium while hydrogen begins to burn in the shell surrounding the core. The star then moves along the Red Giant Branch of the Hertzsprung-Russell (H-R) diagram, with its temperature moderately decreasing, and its radius and luminosity significantly increasing.

As the host star evolves beyond Main Sequence, the orbits of its planets will also evolve. Due to the host's mass loss, its surrounding planets will move outwards (Zahn 1977). On the other hand, tidal interactions tend to shrink the orbital radius of the planets (Villaver et al. 2014). In particular, Villaver et al. 2014 predicted that tidal interaction would cause planets to plunge into the star, and get *engulfed*, before $a/R_s < 3$, where $a$ is the orbital semi-major axis of the planet and $R_s$ is the stellar radius.

The first aim of our paper is to study the distribution of planets around Red Giants. Previous research (Jiang and Zhu 2018) found a power law relation between planet mass and stellar radius. Using the data of newfound exoplanets, we update these results with more extensive analysis. The associated distribution of three variables is focused upon: the mass of the planet ($M_p$), the radius of the star ($R_s$), and the orbital semi-major axis ($a$) — with the aim of gaining additional insight into the evolution of planets as the host star evolves post-Main Sequence.

To scientists, and the general public alike, habitability and the existence of extraterrestrial life is a topic of high interest (Kaltenegger 2017). A habitable zone is an annular region around a given star where any hosted planets have a relatively high likelihood of moderate average surface temperature, allowing for biological life (as we know it) to possibly exist. The HZ is usually determined primarily by the stellar energy flux from the host. More specifically, however, for a planet to be habitable it must not only be in its host's HZ, but also possess the appropriate atmospheric and geological conditions accommodative to maintaining surface liquid water.

It has been predicted by many authors that as the Sun enters the RGB, Earth will no longer be in the Solar System's HZ. As investigated by many studies (Danchi et al. 2005, Lopez et al. 2005, Cuntz et al. 2012, Ramirez and Kaltenegger 2016, Gallet et al. 2019, Sparrman 2022), post-Main Sequence evolution of the Sun will alter its HZ, possibly rendering some of the outer planets' moons habitable to life such as found on Earth.

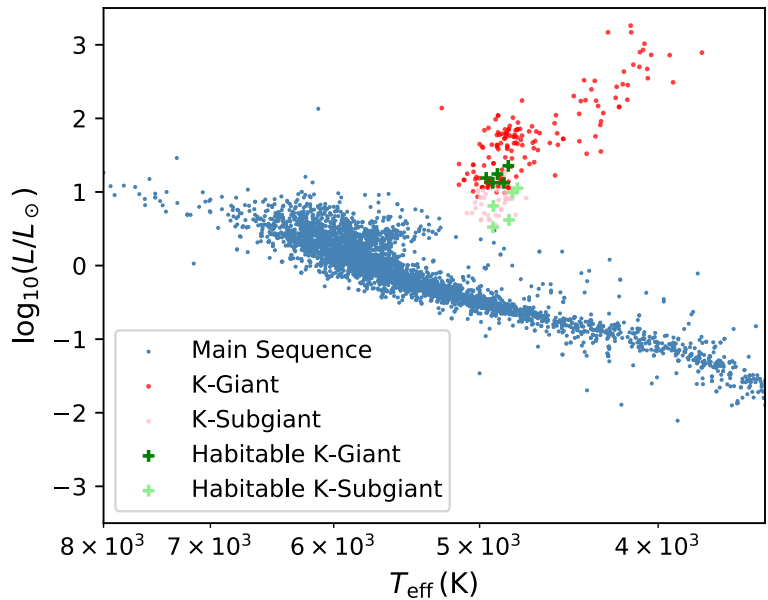

**Figure 1.** H-R diagram of host stars from the 5063 confirmed planets in the NASA Exoplanet Archive. Separated out are the 215 Red Giants via the star's location on the H-R diagram. More specifically, K-giants (with absolute magnitude less than 2.5) are shown in red dots, while K-subgiants (with absolute magnitude between 2.5 and 4) are shown in pink dots. Giants (subgiants) having optimistically calculated habitable planets are labeled by green (light green) plus signs (see Sec. 3 for details).

For a grid of stars with various mass and metallicity, Ramirez and Kaltenegger (2016) explored the evolution models of planets with their stars, and subsequent durations of planets in the HZ in detail. Their findings suggest three candidate systems that will become habitable once the host star becomes a Red Giant. In this paper we apply the criterion used by two previous studies (Ramirez and Kaltenegger 2016, Sparrman 2022), initially proposed by Kopparapu and colleagues (Kopparapu et al 2013), to current data in the NASA Exoplanet Archive (NEA), identifying those exoplanets in the HZ and discussing further parameterized regions not yet observed which may also contain habitable planets.

## 2. Data Collection and Distribution of Planets Around Red Giants

In this section, we briefly introduce our data collection and then discuss the distribution of RG planets in the ($M_p$, $a$, $R_s$) parameter space. In Figure 1, we plot an H-R diagram of host stars using luminosity relative to the Sun ($L/L_\odot$) and stellar surface effective temperature ($T_{eff}$) values from the NEA, identifying 215 Red Giants (and sub-Giants) - plotted in red (and pink) dots. More specifically, we identified these giants according to their location in the H-R Diagram in Figure 1, the giants here are K-giants with absolute magnitudes below 2.5, while subgiants are K-giants with absolute magnitudes between 2.5 and 4. In the Appendix, in Tables 3 and 4, we list the identifiers, stellar mass, stellar radius, orbital semi-major axis and planet mass.

We note that some of these planets, including 42 Dra, γ Dra (Döllinger and Hartmann 2021, henceforth referred to as D&H), and α Tau (Reicher et al. 2019), have been questioned as false positives. D&H further speculated that a substantial fraction of planets around K-giants with radii greater than $21 R_\odot$ can be false positives, based on the congregation of their orbital periods, lack of planet-metallicity correlation, as well as the excess number of planets around K-giants compared with main-sequence stars. We shall make comparisons with D&H in Section 2.3 below.

### 2.1 Planet-Mass-Stellar-Radius Relation

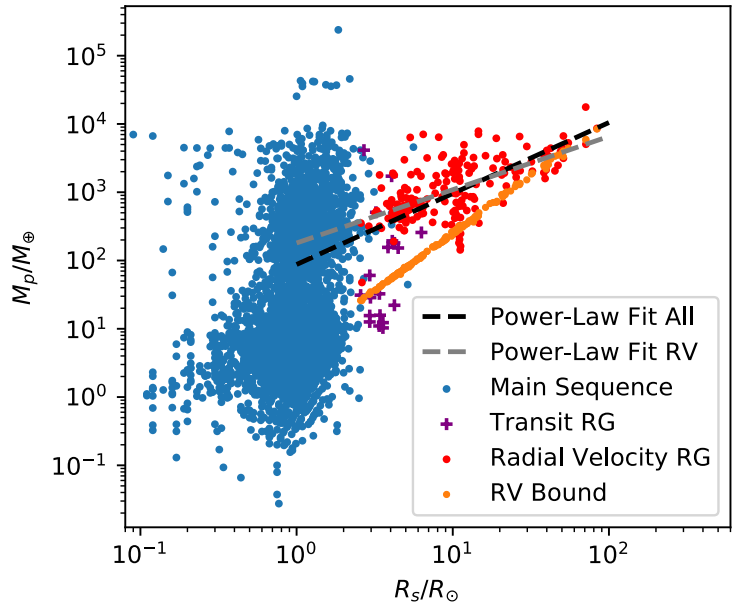

**Figure 2.** $M_p$ vs. $R_s$ plot for MS (blue) and RG hosted planets discovered by radial velocity (red) and transit (purple +) methods. The black dashed line indicates the updated power-law $M_p$ vs. $R_s$ relation for all 215; the gray dashed line corresponds to the power-law fit for planets discovered by radial velocity. Orange dots represent minimum $M_p$ for each red giant that can lead to RV amplitude greater than the stellar intrinsic level obtained by Hekker et al. 2008 [Cf. Eq. (3)].

A previous study (Jiang and Zhu 2018) derived a planet mass-stellar radius relation for 150 exoplanets orbiting Red Giants:

$$M_p/M_\oplus = a(R/R_\odot)^b \quad (1)$$

with best-fit parameters $a$=150 and $b$=0.88. They further argued that Equation 1 is not due to observational bias from the radial-velocity detection method. Folding in the Archive's new data as well, we updated the relation and found a comparably similar result: $a = 87^{+20}_{-16}$ and $b = 1.04 \pm 0.09$, as shown in Figure 2. The adjustment to a lower *a* value and higher *b* value obtained here can most likely be attributed to the post-2018 data points, which have lower values for both planetary mass and stellar radius. When restricted to planets discovered by the radial velocity (RV) method (205 out of the 215 planets), we obtain an alternative fit of $a = 181^{+33}_{-28}$ and $b = 0.78 \pm 0.07$.

**2.2 Observed Evolution of Exoplanet Population as the Host Star Evolves**

Further investigation of the origin of the $M_p$ vs. $R_s$ relation notes that the stellar radius tracks with the post-Main-Sequence evolution stage of the host star. The fact that $M_p$ increases with $R_s$ corresponds to a relative lack of less massive planets around more evolved stars. We shall use Figures 3 and 4, in addition to Figure 2, to further investigate the evolution of the population of exoplanets around stars as they evolve.

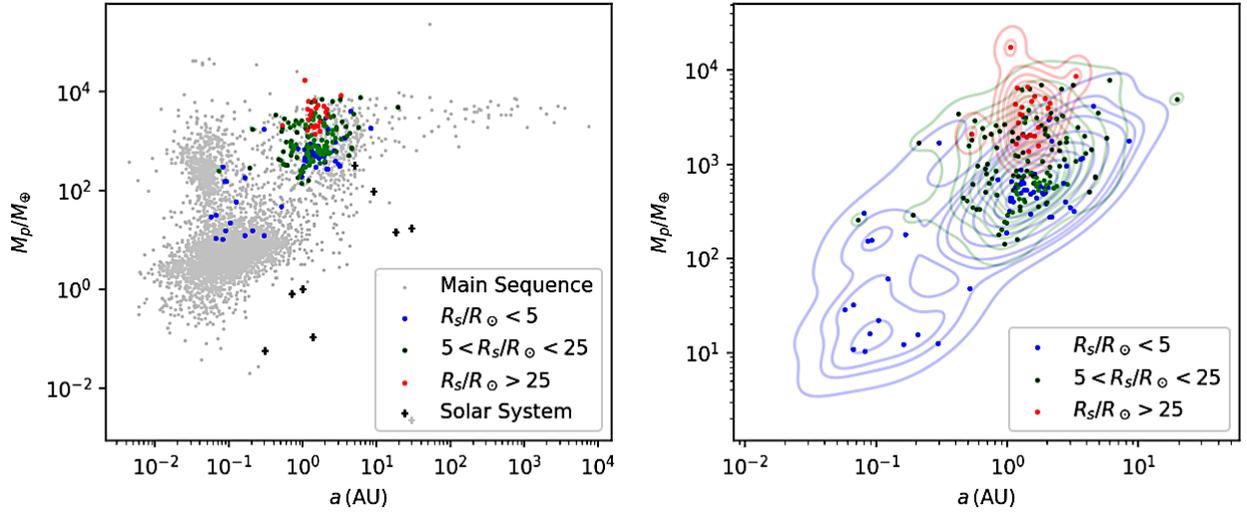

**Figure 3a, 3b.** Left panel (a): $M_p$ vs. *a* plot for main-sequence (silver), Red Giant planets (blue for $R_s/R_\odot < 5$, green for $5 < R_s/R_\odot < 25$, and red for $R_s/R_\odot > 25$), as well as Solar System planets (black). Right panel (b): zoomed-in version for Red Giant planets, with Kernel Density Estimate contours also shown.

In Figure 3a and 3b, we split $R_s$ into three different intervals and plot Main Sequence (silver dots) and Red Giant planets in each interval separately as planet mass $M_p$ (in Earth masses) vs. orbital semi-major axis *a* (in astronomical units). In particular, we separate RG planets into three categories according to $R_s$: $R_s/R_\odot < 5$ (blue dots), $5 < R_s/R_\odot < 25$ (green dots), and $R_s/R_\odot > 25$ (red dots). The $(a, M_p)$ region occupied by RG planets shrinks as $R_s$ increases — from its left side, with small *a*, from the bottom side, with low $M_p$, and from the right side, with large *a*. This shrinkage is best viewed from the right (b) panel of Figure 3, which focuses on the specific region of RG planets and adds contours generated via Kernel Density Estimate (KDE) for clarity. In Figure 4, we plot the orbital semi-major axis vs. stellar radius ratioed to solar radii, illustrating that as $R_s$ increases, the distribution of *a* narrows. In particular, $a/R_s=3$ is seen as a cutoff at lower values of *a*.

At this stage, it is useful to point out that Figure 2 and Figure 4 each separately illustrates the continuous evolutions of $M_p$ and *a*, respectively, as $R_s$ increases, while in Figure 3b we separate the evolution of $R_s$ into three bins and illustrate the lumped ($a$, $M_p$) distribution in each bin. The shrinkage of ($a$,$M_p$) distribution as we progress from blue to green and to red, are more continuously represented as marginal distributions in Figure 4 for $a$ and in Figure 2 for $M_p$.

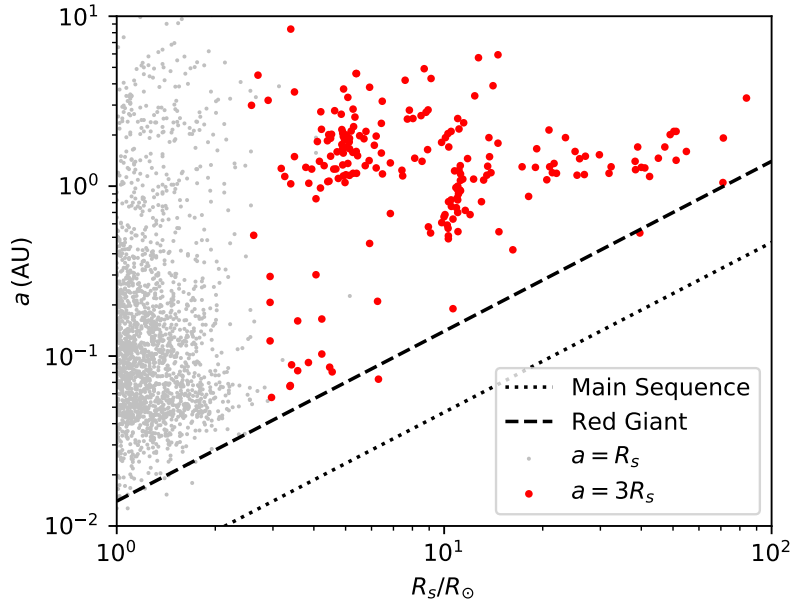

**Figure 4.** $a$ vs. $R_s$ plot for exoplanets around Red Giants (red dots), with Main Sequence planets shown in silver. Solid line indicates $a$=3$R_s$, while dashed line indicates $a$=$R_s$.

With Figures 2, 3 and 4, let us examine the evolution of ($a$, $M_p$) in more detail. Regarding the disappearance of low-mass planets with increasing $R_s$, we can see from Figure 3a (left panel) that for stars with a radius less than 25$R_\odot$ many planets with masses 200 to 1000$M_\oplus$ exist at distances 2 to 3 AU. Yet, these planets are not seen orbiting stars with $R_s$ > 25$R_\odot$ - even though much more massive planets are seen at the same distance. This disappearance of low-mass planets with increasing $R_s$ corresponds directly from the $M_p$ vs. $R_s$ power-law fit in Figure 2. Note that Solar System planets lie on the lower part of the plot; only Jupiter is near the reach of current detection methods. However, Jovian mass exoplanets and comparable orbital distance (~5 AU) are not seen around Red Giants with $R_s$/$R_\odot$ > 25.

For the disappearance of high semi-major axis planets as the star evolves, as can also be seen from Figure 4, one apparent explanation will be the inward migration of hosted planets. Taken a step further, if low-mass planets migrate inward more efficiently migration may explain their disappearance as well. However, it remains unclear whether migration can be sufficiently substantial within the lifetimes of these stars. Another possibility may be that more evolved host stars in our data tend to have lower metallicity and are older aged, and therefore were apt to have differently characterized populations of planets formed around them. However, such differences will likely have to be very substantial to be influential in this respect.

For disappearance of planets with low semi-major axis, it is straightforward to anticipate planets with small orbital distance values to be engulfed and consumed as their host evolves and expands. According to Villaver et al. 2014, tidal interactions tend to speed up the engulfment of planets, and no planets should survive once $a/R_s$ < 3. In Figure 4, we plot the orbital semi-major axis vs. stellar radius ratioed to solar radii, clearly illustrating that $a/R_s=3$ is a cutoff and providing empirical evidence for tidally-accelerated engulfment.

**2.3 Further Interpretations of the Evolutions in Population**

D&H pointed out a concentration of orbital periods (between 300 days and 800 days) for exoplanets around red giants with $R_s > 21 R_\odot$. This is consistent with our Figure 4 and Figure 3b (the horizontal spread of red dots), since orbital period is highly correlated with orbital semi-major axis $a$ and because the host stars have very similar masses. They argued that since the range of period falls within the period of intrinsic variations of stars (as modeled by Saio et al. 2015), and hence a fraction of these may not be actual planets. On the other hand, they provided plausible reasons for planets outside of this period range not to be discovered. For longer periods (corresponding to larger $a$), this could be due to the smaller RV variation being hidden under intrinsic fluctuations of the surface of the host star, while for shorter periods (corresponding to smaller $a$), this could be due to the engulfment of planets by their host stars.

In this way, one might speculate that the population of exoplanets with the concentrated distribution of orbital periods can still exist, but some data points are potentially contaminated with intrinsic stellar oscillations. We shall contrast our results with those of D&H, and extend the discussion.

D&H argued that planets with periods shorter than 300 days could be absent due to engulfment, similar to what we propose in the previous section. We note that these planets are the closest to the star, tend to cause the highest RV variations, and are hence *the least* prone to be missed in RV measurements. In Figure 4, where we plot semimajor axis $a$ and $R_s$, when combined with the results of Villaver et al. 2014, provides direct evidence that short-period exoplanets are missing due to engulfment.

For longer periods than 800 days, D&H argued that they may be missing due to the fact that they cause lower variation in RV and hence are more prone to be hidden below intrinsic variations of the host stars. We note that RV is a combined effect between planet mass and semi-major axis, with amplitude $K_1$ of RV given by:

$$K_1 = \left(\frac{2\pi G}{P}\right)^{1/3} \frac{M_p \sin i}{\sqrt{1-e^2} M_s^{2/3}} \propto M_p a^{-1/2}, \qquad (2)$$

with $P$ the orbital period, $i$ the orbital inclination angle and $e$ the orbital eccentricity. Here we have highlighted its dependence on $M_p$ and $a$. In this way, the disappearance of low-mass and large $a$ planets as $R_s$ increases could both be due to an observational cutoff in $K_1$.

Under this assumption, we can first make a re-interpretation of the $M_p$-$R_s$ relation seen in Figure 2. More specifically, we use results from Hekker et al. 2008, who noticed that for stars with lower surface gravity $g$ (i.e., larger radii), their amplitudes of RV variations tend to increase. They found a baseline value, given approximately by

$$K_1^{\text{int}} = 2 \times 10^3 \ [g/(\text{cm/s}^2)]^{-0.6} \text{m/s}, \qquad (3)$$

which they *interpret* as arising from intrinsic fluctuations of the star. Here, $g$ is the surface gravitational acceleration of the star. For each red giant, combining Eqs. (2) and (3), assuming $e=0$, and using $a = 3R_s$, we obtain the minimum planet mass $M_p^{\text{min}}$ the star can host, this in order for the $K_1$ due to the planet to be greater than the intrinsic $K_1^{\text{int}}$:

$$M_\text{p}^\text{min} = \sqrt{\frac{3M_s R_s K_1^\text{int}}{G}}. \tag{4}$$

$M_\text{p}^\text{min}$ is plotted as orange dots in Figure 2; indeed providing an excellent lower bound for planets around substantially evolved stars. We also note that in Figures 2 and 3b the upper limit in $M_\text{p}$ for higher radii stars are the same as for lower radii stars.

Together, the varying lower bound and roughly constant higher bound in $M_\text{p}$ for increasing $R_s$ is consistent with the notion that the mass distribution of exoplanets stay the same as the star evolves, yet the difficulty in observation gradually eliminates low-mass planets from the observed distribution. Further independent studies into the amplitude of stellar oscillations and practical limitations of the RV method are necessary to confirm this picture.

## 3. Habitable Planets Around Red Giants

In this section we discuss the habitability of planets around Red Giants, briefly reviewing habitability criteria in 3.1 and presenting our findings in 3.2.

### 3.1 Criteria for Habitability

There exist multiple habitability conditions for a given exoplanet (or exomoon); most of which rely on the existence of water in liquid form to be present on at least a portion of that world's surface. The simplest criterion uses equilibrium temperature, namely, the black-body radiation from the planet has to balance the radiation it absorbs from the star. If we define S as the flux of radiation from the host, this given by

$$S = \frac{L_s}{4\pi a^2} \tag{5}$$

where $L_s$ is the star's luminosity and $a$ is the orbital semi-major axis of the star's exoplanet, the equilibrium temperature of the exoplanet is then given by

$$T_\text{eq} = k \left[\frac{S(1-A)}{4\sigma}\right]^{1/4} \tag{6}$$

where $A$ is the planetary albedo and $\sigma$ is the Stefan-Boltzmann constant. The simplest habitability condition is $273\text{ K} < T_\text{eq} < 373\text{ K}$, with the low $T_\text{eq}$ defining the Outer boundary of the Habitable Zone (OHZ) and the high $T_\text{eq}$ defining the Inner boundary of the Habitable Zone (IHZ). The scaler quantity k is a correction factor that can be used to approximately incorporate the greenhouse effect of an assumed planetary atmosphere.

More realistic criteria exist in the literature. In this paper, we shall adopt two criteria obtained by previous study (Kopparapu et al. 2013) in which an *effective* solar flux is expressed in terms of

$$S_\text{eff} \equiv S/S_\oplus \tag{7}$$

where $S_\oplus$ is the current solar energy flux at the location of the Earth, as well as the temperature $T$ of the host star. Note that $S_\text{eff}$ is dimensionless. In this paper, we shall adopt two different ways to define HZ boundaries, one conservative, the other optimistic. The conservative HZ accounts for greenhouse effects in the atmosphere of the planet, taking the inner boundary to be defined by the moist greenhouse effect where $S_\text{eff}$ allows sufficient water vapor to exist in the stratosphere. The outer boundary is defined by the maximum heat retained by the planet while still providing habitable conditions. This is also known as the maximum greenhouse effect.

We also summarized the boundaries using the following fitting formula (Kopparapu et al. 2013) for the host star temperature range of 2600 K < $T$ < 7200 K:

$$S_{\text{limit}}(T) = S_0 + aT_* + bT_*^2 + cT_*^3 + dT_*^4, \quad T_* = T - 5780\,\text{K}, \tag{8}$$

where values of $a$, $b$, $c$ and $d$ for conservative/optimistic, inner/outer boundaries are reproduced in Table 1.

**Table 1.** Fitting parameters $S_0$, $a$, $b$, $c$ and $d$ adapted from previous study (Kopparapu et al. 2013)

|  | $S_0$ | $a$ | $b$ | $c$ | $d$ |
|---|---|---|---|---|---|
| **Recent Venus** (optimistic inner boundary) | 1.7753 | 1.4316×10⁻⁴ | 2.9875×10⁻⁹ | −7.5702×10⁻¹² | −1.1635×10⁻¹⁵ |
| **Moist Green House** (conservative inner boundary) | 1.0140 | 8.1774×10⁻⁵ | 1.7063×10⁻⁹ | −4.3241×10⁻¹² | −6.6462×10⁻¹⁶ |
| **Maximum Green House** (conservative outer boundary) | 0.3438 | 5.8942×10⁻⁵ | 1.6558×10⁻⁹ | −3.0045×10⁻¹² | −5.2983×10⁻¹⁶ |
| **Early Mars** (optimistic outer boundary) | 0.3179 | 5.4513×10⁻⁵ | 1.5313×10⁻⁹ | −2.7786×10⁻¹² | −4.8997×10⁻¹⁶ |

**Table 2.** Conservative (shaded) and optimistically (unshaded) habitable planets using the Kopparapu et al. 2013 criterion. Note that most HZ planets are orbiting subgiants (with magnitude between 2.5 and 4).

| Planet Name | Discovery Paper | Spectral Type (NEA) | Abs Mag | Host Mass (M) | Host Radius (R) | Orbital Period (days) | $S_{\text{eff}}$ | Planet Mass ($M_J$) |
|---|---|---|---|---|---|---|---|---|
| HD 1605 c | Hirakawa et al. 2015 | K1 IV | 2.8 | 1.33 | 3.49 | 233 | 0.50 | 3.62 |
| HD 219415 b | Gettel et al. 2012 | K0 III | 2.8 | 1.0 | 2.90 | 207 | 0.41 | 1.0 |
| HD 4732 c | Sato et al. 2013 | K0 IV | 2.2 | 1.74 | 5.40 | 118 | 0.73 | 2.37 |
| HD 73534 b | Valenti et al. 2009 | G5 (IV) | 3.6 | 1.16 | 2.58 | 168 | 0.37 | 1.11 |
| HIP 56640 b | Jones et al. 2021 | K1 III | 2.5 | 1.04 | 4.93 | 157 | 0.81 | 3.67 |
| HD 125390 b | Luhn et al. 2019 | G7 V (III) | 2.3 | 1.36 | 6.47 | 342 | 1.33 | 22.2 |
| HD 145934 b | Feng et al. 2015 | K0 | 1.7 | 1.75 | 5.38 | 215 | 1.07 | 2.28 |
| HD 94834 b | Luhn et al. 2019 | K0 | 2.6 | 1.11 | 4.20 | 38 | 1.31 | 1.26 |
| HD 95089 c | Bryan et al. 2016 | G8/K0 IV | 2.23 | 1.54 | 5.08 | 66 | 1.20 | 3.45 |
| HIP 67851 c | Jones et al. 2015 | K0 III | 2.14 | 1.63 | 5.92 | 167 | 1.20 | 6.3 |

As noted in Table 1, a more optimistic approach uses the (theorized) history of Solar System planets Venus and Mars to determine the inner and outer bounds of the HZ. Here, the inner boundary of the HZ is based on the assertion that Venus has not had liquid water on its surface for only the past billion years – i.e., a billion years ago (recent) Venus might have had surface conditions suitable for water to exist. On the other hand, there is mounting evidence that (early) Mars had liquid water flowing on its surface 3.8 billion years ago. For these reasons, they define the inner boundary using the $S_{\text{eff}}$ of recent Venus and the outer boundary using the $S_{\text{eff}}$ of early Mars.

In Table 2, we list conservative and optimistic habitable zone RG-hosted planets. All planets are gas giants with masses ranging from 1 to 22 Jupiter masses ($M_J$). It should be noted for HD 73534 the luminosity class was missing from the archive and is backfilled here from other sourcing as IV (indicating subgiant). As well for HD 125390, its luminosity was mistyped as V (dwarf) but has been correctly listed in Table 2 as III (giant). On the fourth column, we list the absolute magnitude computed from the V-magnitude and distance data from the Archive, noting that the classification from absolute magnitude does not always agree with the third column. The Red Giant hosts of planets in Table 2 are also shown as green dots in the H-R diagram of Figure 1. As can be readily perceived from Table 2 and Figure 1, these host stars are all in their early stages of evolution on the RGB.

### 3.2 Red Giant Planets in the Habitable Zones

From the NASA Exoplanet Archive, we collected values for stellar luminosity and orbital semi-major axis to calculate $S_{eff}$. In Figure 5, we show the Red Giant planets on the $T_{eff}$ vs. $S_{eff}$ plot with lines indicating HZ boundaries. From the plot, it can be seen that there is a substantial difference between boundaries for the $T_{eq}$ HZ and Kopparapu HZ. We highlighted Kopparapu et al. 2013 optimistically habitable planets in green.

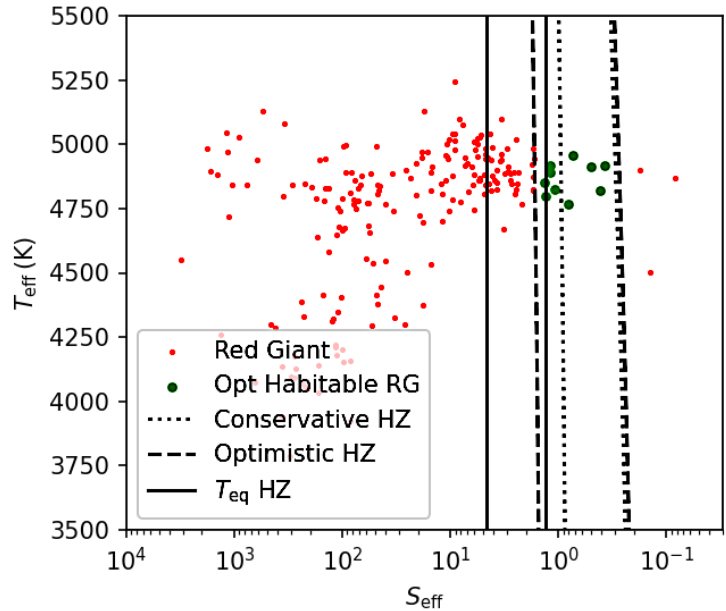

**Figure 5.** $T_{eff}$ vs. $S_{eff}$ plot for Red Giant planets (red dots) with optimistic habitable planets denoted in green. Boundaries for $T_{eq}$, conservative and optimistic HZ are shown in solid, dotted and dashed lines, respectively.

Figure 6 shows Red Giant and Main Sequence planets on a semi-major axis vs. stellar radius plot with habitable planets indicated (light green dots for habitable planets around MS, and dark green dots for habitable planets around subgiants and giants). We also indicate, with purple line segments, the optimistic HZ of the host stars of all planets around subgiants and giants. As illustrated, habitable planets --- and indeed habitable zones --- follow a track with $a$ increasing as $R_s$ increases, this attributable to stars with larger radii - and thus greater luminosity - having HZs farther out. We also see that the track stops at $R_s \sim 8R_\odot$, far below the maximum $R_s$ of Red Giants, clearly suggesting there is a missing population of planets around Red Giants with $a$ above ~4 AU.

This also explains why most of the conservative habitable planets so far discovered are orbiting subgiants, or giants at their early stages of evolution. If there are more Red Giant planets in this undetected region, it is likely there are undiscovered habitable zone planets as well. Here, we also recall that even though outer planets of our Solar System can become habitable as the Sun evolves, these planets are far from the detectable zone, as we can see from Figure 3a.

As with Main Sequence hosts, these large semi-major axis planets were all discovered by direct imaging. However, using the same detection method to find similar planets around Red Giants may be difficult due to the direct imaging method disfavoring systems with large radii hosts.

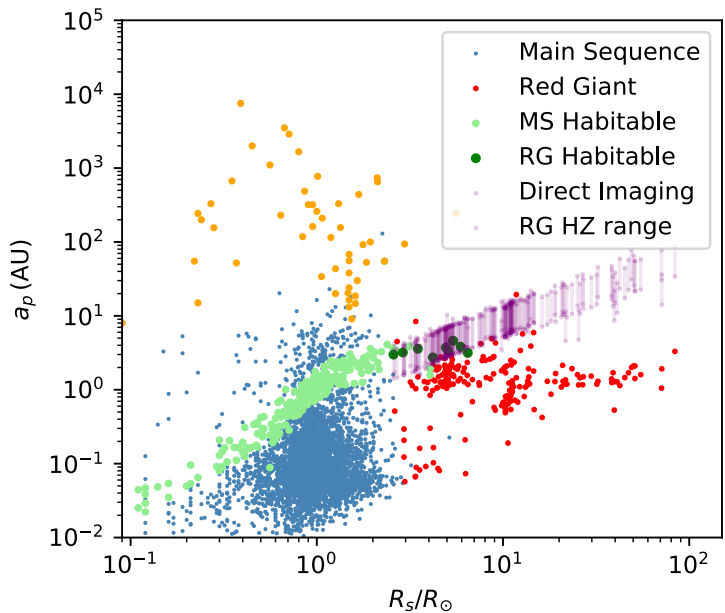

**Figure 6.** Semi-major axis *a* vs. stellar radius $R_s$ plot of Red Giant (red) and Main Sequence planets (blue) with optimistically habitable planets in green (light green for Main Sequence and darker green for Red Giant planets.) Planets discovered by direct imaging are shown in orange. With purple vertical line segments, we indicate the optimistic HZ of each giant.

**4. Conclusions and Discussions**

In this paper we take new data from NASA's Exoplanet Archive to update and further investigate trends regarding Red Giant systems. First, we revisit the Planet Mass-Stellar Radius relation previously found (Jiang and Zhu 2018), observing that a similar power-law relationship is bolstered with the addition of more than 50 new data points. However, in our results we noted a steeper power-law relation due to the additional data points with lower mass and stellar radius values. As we focus on planets discovered by the radial velocity technique (205 out of 215 planets), the steepness of the power-law was attenuated. To further explore this trend, we separate Red Giant hosted exoplanets according to the radii of their hosts and plot planet mass against semi-major axis (Figure 3). As stellar radius increases, the region occupied by planets in the graph shrinks and for planets with smaller orbital semi-major axes, we found their disappearance to be consistent with tidal engulfment of planets where $a/R_s < 3$ (Figure 4).

For the disappearance of planets with lower masses and those with larger orbital semi-major axes, we did not find compelling astrophysical reasons; this disappearance could be due to observational selection effects of the radial velocity method used to discover the vast majority of planets in these regions. Since lower mass and larger orbital semi-major axis correspond to lower amplitudes of radial velocity, the disappearance can be attributed to a higher detection threshold for the amplitude of radial velocity oscillations among more evolved Red Giants. We showed that in order for this selection effect to be the origin of such disappearance, the level of intrinsic RV fluctuation of Red Giants should depend on surface gravity following equation (3), which was proposed by Hekker et al. 2018. However, selection effects may also arise due to eccentricity and stellar mass. As this possibility is beyond the scope of this paper, further investigations of the origin of such selection effects are left to future studies.

Next, we examine the habitability of Red Giant exoplanets. To determine the habitable zone, we adopt criteria proposed by Kopparapu et al. 2013 and with this method found ten planets in the optimistic HZ, five of which are in the conservatively calculated HZ. However, all of these planets are gas giants and, therefore, very likely uninhabitable by life as we presently know it. Nevertheless, these planets may themselves host habitable exomoons.

Finally, with habitable zone exoplanets identified, we look at possible detection bias. We see that their orbital semi-major axis increases with stellar radii until $R_s/R_\odot \sim 7$. However, this does not necessarily rule out further habitable zone exoplanets and it is very likely there are more HZ Red Giant exoplanets with a semi-major axis greater than ~4 AU. Even though some such planets can be seen around Main Sequence stars via direct imaging, similar planets around Red Giant stars have not yet been found. While the limitations of current imaging methods may preclude detecting planets around Red Giant stars, more advanced instrumentation coming online in the near term may enable this technique to be used for at least some Red Giant hosted exoplanetary systems.

**Acknowledgements**

This research was conducted at the NASA sponsored Jet Propulsion Laboratory, California Institute of Technology (Caltech) and has made use of the NASA Exoplanet Archive, which is operated by the Caltech, under contract with the NASA under the Exoplanet Exploration Program.

**Data Statement:**

The data underlying this article can be downloaded from the NASA exoplanet archive at https://exoplanetarchive.ipac.caltech.edu. The method of data calculation and analysis are fully described in the article.

**Author Contributions:**

Conceptualization, J.H.J.; methodology, J.H.J., R.E.C; software, R.E.C and J.H.J; validation, J.H.J. and Y.C.; formal analysis, R.E.C and J.H.J.; investigation, R.E.C., J.H.J. and P.E.R.; resources, J.H.J.; data curation, R.E.C and J.H.J.; writing—original draft preparation, R.E.C.; writing—review and editing, J.H.J., P.E.R., and K.A.F; visualization, R.E.C and J.H.J; supervision, J.H.J. and Y.C.; project administration, J.H.J.; funding acquisition, J.H.J. All authors have read and agreed to the published version of the manuscript.

**Competing Interest:**

Authors declare no competing interest.

**Appendix. Exoplanets around K-Giants and K-Subgiants from the NASA Exoplanet Archive**

Table 3. Exoplanets hosted by K-Giants from the NASA Exoplanet Archive (a total of 169 planets)

| Planet Name | $M_s$ ($M_\odot$) | $R_s$ ($R_\odot$) | $a$ (AU) | $M_p$ ($M_J$) | Planet Name | $M_s$ ($M_\odot$) | $R_s$ ($R_\odot$) | $a$ (AU) | $M_p$ ($M_J$) | Planet Name | $M_s$ ($M_\odot$) | $R_s$ ($R_\odot$) | $a$ (AU) | $M_p$ ($M_J$) |
|---|---|---|---|---|---|---|---|---|---|---|---|---|---|---|
| 11 Com b | 2.70 | 19.00 | 1.29 | 19.40 | HD 161178 b | 1.06 | 10.95 | 0.85 | 0.57 | HD 81817 b | 4.30 | 83.80 | 3.30 | 27.10 |
| 11 UMi b | 2.78 | 29.79 | 1.53 | 14.74 | HD 1690 b | 1.86 | 21.66 | 1.36 | 8.79 | HD 82886 b | 2.53 | 5.26 | 1.58 | 2.33 |
| 14 And b | 2.20 | 11.00 | 0.83 | 4.80 | HD 17092 b | 6.73 | 13.58 | 1.31 | 10.13 | HD 86950 b | 1.66 | 8.80 | 2.72 | 3.60 |
| 17 Sco b | 1.22 | 25.92 | 1.45 | 4.32 | HD 173416 b | 2.00 | 13.50 | 1.16 | 2.70 | HD 95089 b | 1.54 | 5.08 | 1.36 | 1.26 |
| 18 Del b | 2.30 | 8.50 | 2.60 | 10.30 | HD 175541 b | 1.39 | 4.19 | 0.98 | 0.60 | HD 95089 c | 1.54 | 5.08 | 3.33 | 3.45 |
| 24 Boo b | 0.99 | 10.64 | 0.19 | 0.91 | HD 180053 b | 1.75 | 4.06 | 0.84 | 2.19 | HD 95127 b | 3.70 | 41.01 | 1.28 | 10.63 |
| 24 Sex b | 1.54 | 4.90 | 1.33 | 1.99 | HD 180314 b | 2.20 | 8.13 | 1.46 | 20.13 | HD 96063 b | 1.37 | 4.75 | 1.11 | 1.27 |
| 24 Sex c | 1.54 | 4.90 | 2.08 | 0.86 | HD 181342 b | 1.69 | 4.71 | 1.59 | 2.54 | HD 96127 b | 10.94 | 51.10 | 1.42 | 20.96 |
| 4 UMa b | 1.23 | 18.11 | 0.87 | 7.10 | HD 18742 b | 1.36 | 5.13 | 1.82 | 3.40 | HD 96992 b | 0.96 | 7.43 | 1.24 | 1.14 |
| 42 Dra b | 0.98 | 22.03 | 1.19 | 3.88 | HD 192699 b | 1.38 | 4.41 | 1.06 | 2.10 | HD 99283 b | 1.76 | 11.21 | 1.08 | 0.97 |
| 6 Lyn b | 1.44 | 5.20 | 2.11 | 2.01 | HD 200964 b | 1.39 | 4.92 | 1.57 | 1.60 | HD 99706 b | 1.46 | 5.52 | 1.98 | 1.23 |
| 7 CMa b | 1.34 | 4.87 | 1.76 | 1.85 | HD 200964 c | 1.39 | 4.92 | 1.96 | 1.21 | HD 99706 c | 1.72 | 5.40 | | 5.69 |
| 7 CMa c | 1.34 | 4.87 | 2.15 | 0.87 | HD 202696 b | 1.91 | 6.43 | 1.57 | 2.00 | HIP 105854 b | 2.10 | 10.31 | 0.81 | 8.20 |
| 75 Cet b | 2.49 | 10.50 | 2.10 | 3.00 | HD 202696 c | 1.91 | 6.43 | 2.34 | 1.86 | HIP 107773 b | 2.42 | 11.60 | 0.72 | 1.98 |
| 8 UMi b | 1.44 | 10.30 | 0.49 | 1.31 | HD 208527 b | 1.60 | 51.10 | 2.10 | 9.90 | HIP 114933 b | 1.39 | 5.27 | 2.84 | 1.94 |
| 81 Cet b | 2.40 | 11.00 | 2.50 | 5.30 | HD 208897 b | 1.25 | 4.98 | 1.05 | 1.40 | HIP 63242 b | 1.54 | 10.28 | 0.57 | 9.18 |
| 91 Aqr b | 1.40 | 11.00 | 0.70 | 3.20 | HD 210702 b | 1.61 | 4.92 | 1.15 | 1.81 | HIP 65891 b | 2.50 | 8.93 | 2.81 | 6.00 |
| BD+03 2562 b | 1.14 | 32.35 | 1.30 | 6.40 | HD 212771 b | 1.56 | 5.27 | 1.19 | 2.39 | HIP 67537 b | 2.41 | 8.69 | 4.91 | 11.10 |
| BD+15 2375 b | 1.08 | 8.95 | 0.58 | 1.06 | HD 216536 b | 0.81 | 9.83 | 0.61 | 1.05 | HIP 67851 b | 1.63 | 5.92 | 0.46 | 1.38 |
| BD+15 2940 b | 1.10 | 14.70 | 0.54 | 1.11 | HD 219139 b | 1.46 | 11.22 | 0.94 | 0.78 | HIP 67851 c | 1.63 | 5.92 | 3.82 | 6.30 |
| BD+20 2457 b | 10.83 | 71.02 | 1.05 | 55.59 | HD 220074 b | 2.20 | 54.92 | 1.60 | 16.64 | HIP 8541 b | 1.17 | 7.83 | 2.80 | 5.50 |
| BD+20 2457 c | 2.80 | 49.00 | 2.01 | 12.47 | HD 22532 b | 1.57 | 5.69 | 1.90 | 2.12 | HIP 97233 b | 1.93 | 5.34 | 2.55 | 20.00 |
| BD+20 274 b | 0.80 | 17.30 | 1.30 | 4.20 | HD 233604 b | 1.50 | 10.90 | 0.75 | 6.58 | IC 4651 9122 b | 2.10 | 10.27 | 2.04 | 6.30 |
| BD+48 738 b | 0.74 | 11.00 | 1.00 | 0.91 | HD 238914 b | 1.47 | 12.73 | 5.70 | 6.00 | Kepler-91 b | 1.31 | 6.30 | 0.07 | 0.81 |
| BD+48 740 b | 1.09 | 10.33 | 1.70 | 1.70 | HD 240210 b | 0.82 | 25.46 | 1.16 | 5.21 | NGC 2682 Sand 364 b | 9.06 | 39.59 | 0.53 | 6.69 |

| Name | | | | | Name | | | | | Name | | | | |
|---|---|---|---|---|---|---|---|---|---|---|---|---|---|---|
| BD+49 828 b | 1.52 | 7.60 | 4.20 | 1.60 | HD 240237 b | 8.76 | 71.23 | 1.92 | 15.89 | NGC 2682 Sand 978 b | 1.37 | 21.02 | | 2.18 |
| BD-13 2130 b | 2.12 | 19.17 | 1.66 | 9.78 | HD 24064 b | 1.61 | 40.00 | 1.29 | 12.89 | TYC 0434-04538-1 b | 1.04 | 9.99 | 0.66 | 6.10 |
| HD 100655 b | 2.28 | 10.06 | 0.68 | 1.61 | HD 25723 b | 2.12 | 13.76 | 1.49 | 2.50 | TYC 1422-614-1 b | 1.15 | 6.85 | 0.69 | 2.50 |
| HD 102272 b | 1.45 | 10.30 | 0.51 | 4.94 | HD 28678 b | 1.53 | 6.48 | 1.18 | 1.54 | TYC 1422-614-1 c | 1.15 | 6.85 | 1.37 | 10.00 |
| HD 102329 b | 3.21 | 9.82 | 1.81 | 8.16 | HD 2952 b | 1.97 | 10.76 | 1.23 | 1.37 | TYC 3318-01333-1 b | 1.19 | 5.90 | 1.41 | 3.42 |
| HD 102329 c | 1.30 | 6.30 | | 1.52 | HD 30856 b | 1.17 | 4.40 | 1.85 | 1.55 | TYC 3667-1280-1 b | 1.87 | 6.26 | 0.21 | 5.40 |
| HD 102956 b | 1.66 | 4.55 | 0.08 | 0.96 | HD 32518 b | 1.13 | 10.22 | 0.59 | 3.04 | TYC 4282-00605-1 b | 0.97 | 16.21 | 0.42 | 10.78 |
| HD 104985 b | 2.30 | 11.00 | 0.95 | 8.30 | HD 33844 b | 1.84 | 5.39 | 1.60 | 2.01 | alf Ari b | 1.50 | 13.90 | 1.20 | 1.80 |
| HD 108863 b | 1.59 | 5.74 | 1.32 | 2.41 | HD 33844 c | 1.78 | 5.29 | 2.24 | 1.75 | alf Tau b | 1.13 | 45.10 | 1.46 | 6.47 |
| HD 10975 b | 1.41 | 11.16 | 0.95 | 0.45 | HD 360 b | 1.69 | 10.86 | 0.98 | 0.75 | bet Cnc b | 1.70 | 47.20 | 1.70 | 7.80 |
| HD 110014 b | 2.17 | 20.90 | 2.14 | 11.09 | HD 40956 b | 2.00 | 8.56 | 1.40 | 2.70 | bet UMi b | 1.40 | 38.30 | 1.40 | 6.10 |
| HD 111591 b | 1.94 | 8.03 | 2.50 | 4.40 | HD 4313 b | 1.63 | 5.14 | 1.16 | 1.93 | eps CrB b | 1.70 | 21.00 | 1.30 | 6.70 |
| HD 112640 b | 1.80 | 39.00 | 1.70 | 5.00 | HD 4732 b | 1.74 | 5.40 | 1.19 | 2.37 | eps Tau b | 2.70 | 13.70 | 1.93 | 7.60 |
| HD 113996 b | 1.49 | 25.11 | 1.60 | 6.30 | HD 4732 c | 1.74 | 5.40 | 4.60 | 2.37 | gam 1 Leo b | 1.23 | 31.88 | 1.19 | 8.78 |
| HD 116029 b | 0.83 | 4.89 | 1.65 | 1.40 | HD 47366 b | 2.19 | 6.20 | 1.28 | 2.30 | gam Cep b | 1.40 | 4.90 | 2.05 | 9.40 |
| HD 116029 c | 1.33 | 4.60 | | 1.27 | HD 47366 c | 2.19 | 6.20 | 1.97 | 1.88 | gam Lib b | 1.47 | 11.10 | 1.24 | 1.02 |
| HD 11755 b | 0.72 | 20.58 | 1.09 | 5.63 | HD 47536 b | 2.10 | 23.47 | 1.93 | 7.32 | gam Lib c | 1.47 | 11.10 | 2.17 | 4.58 |
| HD 11977 b | 1.91 | 10.09 | 1.93 | 6.54 | HD 4760 b | 1.05 | 42.40 | 1.14 | 13.90 | gam Psc b | 0.99 | 11.20 | 1.32 | 1.34 |
| HD 120084 b | 2.39 | 9.12 | 4.30 | 4.50 | HD 4917 b | 1.32 | 5.01 | 1.17 | 1.62 | iot Dra b | 1.54 | 11.79 | 1.45 | 11.82 |
| HD 125390 b | 1.36 | 6.47 | 3.16 | 22.16 | HD 5583 b | 1.01 | 9.09 | 0.53 | 5.78 | iot Dra c | 1.54 | 11.79 | 19.40 | 15.60 |
| HD 12648 b | 0.67 | 11.02 | 0.54 | 1.96 | HD 5608 b | 1.53 | 5.14 | 1.91 | 1.68 | kap CrB b | 1.50 | 4.85 | 2.65 | 2.00 |
| HD 131496 b | 1.34 | 4.44 | 2.01 | 1.80 | HD 5891 b | 1.93 | 10.64 | 0.64 | 7.63 | mu Leo b | 1.50 | 11.40 | 1.10 | 2.40 |
| HD 13189 b | 2.24 | 38.41 | 1.25 | 10.95 | HD 59686 A b | 1.90 | 13.20 | 1.09 | 6.92 | nu Oph b | 2.70 | 14.60 | 1.79 | 22.21 |
| HD 139357 b | 1.35 | 11.47 | 2.36 | 9.76 | HD 60292 b | 1.70 | 27.00 | 1.50 | 6.50 | nu Oph c | 2.70 | 14.60 | 5.93 | 24.66 |
| HD 14067 b | 2.40 | 12.40 | 3.40 | 7.80 | HD 62509 b | 2.00 | 8.90 | 1.64 | 2.30 | ome Ser b | 2.17 | 12.30 | 1.10 | 1.70 |
| HD 145457 b | 1.23 | 10.52 | 0.76 | 2.23 | HD 64121 b | 1.64 | 5.44 | 1.51 | 2.56 | omi CrB b | 2.13 | 10.50 | 0.83 | 1.50 |
| HD 145934 b | 1.75 | 5.38 | 4.60 | 2.28 | HD 66141 b | 1.10 | 21.40 | 1.20 | 6.00 | omi UMa b | 3.09 | 14.10 | 3.90 | 4.10 |
| HD 14787 b | 1.43 | 5.01 | 1.70 | 1.12 | HD 69123 b | 1.68 | 7.72 | 2.48 | 3.04 | tau Gem b | 2.30 | 26.80 | 1.17 | 20.60 |
| HD 1502 b | 1.46 | 4.67 | 1.26 | 2.75 | HD 72490 b | 1.21 | 4.96 | 1.88 | 1.77 | ups Leo b | 1.48 | 11.22 | 1.18 | 0.51 |
| HD 152581 b | 1.30 | 5.14 | 1.66 | 1.87 | HD 76920 b | 1.17 | 7.47 | 1.15 | 3.93 | xi Aql b | 2.20 | 12.00 | 0.68 | 2.80 |

| Planet Name | $M_s$ ($M_\odot$) | $R_s$ ($R_\odot$) | $a$ (AU) | $M_p$ ($M_J$) |
|---|---|---|---|---|
| HD 155233 b | 1.69 | 5.03 | 2.00 | 2.60 |
| HD 158996 b | 1.80 | 50.30 | 2.10 | 14.00 |
| HD 79181 b | 1.28 | 11.06 | 0.90 | 0.64 |
| HD 81688 b | 2.10 | 13.00 | 0.81 | 2.70 |

**Table 4.** Exoplanets hosted by K-Subgiants from the NASA Exoplanet Archive (a total of 46 planets)

| Planet Name | $M_s$ ($M_\odot$) | $R_s$ ($R_\odot$) | $a$ (AU) | $M_p$ ($M_J$) | Planet Name | $M_s$ ($M_\odot$) | $R_s$ ($R_\odot$) | $a$ (AU) | $M_p$ ($M_J$) | Planet Name | $M_s$ ($M_\odot$) | $R_s$ ($R_\odot$) | $a$ (AU) | $M_p$ ($M_J$) |
|---|---|---|---|---|---|---|---|---|---|---|---|---|---|---|
| EPIC 248847494 b | 0.90 | 2.70 | 4.50 | 13.00 | HD 29399 b | 1.17 | 4.50 | 1.91 | 1.57 | Kepler-1004 b | 1.11 | 3.39 | 0.07 | 0.10 |
| HD 136418 b | 1.48 | 3.78 | 1.29 | 2.14 | HD 33142 b | 1.41 | 4.45 | 1.07 | 1.39 | Kepler-1270 b | 1.28 | 3.38 | 0.07 | 0.03 |
| HD 142245 b | 3.50 | 4.63 | 2.78 | 3.07 | HD 33142 c | 1.62 | 4.14 | | 5.97 | Kepler-278 b | 1.08 | 2.94 | 0.21 | 0.05 |
| HD 148427 b | 1.64 | 3.86 | 1.04 | 1.30 | HD 73534 b | 1.16 | 2.58 | 2.99 | 1.11 | Kepler-278 c | 1.08 | 2.94 | 0.29 | 0.04 |
| HD 158038 b | 1.30 | 4.50 | 1.50 | 1.53 | HD 75784 b | 1.26 | 3.40 | 1.03 | 1.00 | Kepler-391 b | 1.03 | 3.57 | 0.08 | 0.03 |
| HD 1605 b | 1.33 | 3.49 | 1.49 | 0.93 | HD 75784 c | 1.26 | 3.40 | 8.40 | 5.64 | Kepler-391 c | 1.03 | 3.57 | 0.16 | 0.04 |
| HD 1605 c | 1.33 | 3.49 | 3.58 | 3.62 | HD 94834 b | 1.11 | 4.20 | 2.74 | 1.26 | Kepler-432 b | 1.32 | 4.06 | 0.30 | 5.41 |
| HD 167042 b | 1.72 | 4.30 | 1.32 | 1.70 | HD 98219 b | 1.41 | 4.60 | 1.26 | 1.96 | Kepler-432 c | 1.32 | 4.06 | | 2.43 |
| HD 177830 b | 1.70 | 3.26 | 1.14 | 1.69 | HIP 56640 b | 1.04 | 4.93 | 3.73 | 3.67 | Kepler-56 b | 1.32 | 4.23 | 0.10 | 0.07 |
| HD 177830 c | 1.47 | 2.62 | 0.51 | 0.15 | HIP 74890 b | 1.74 | 5.77 | 2.10 | 2.40 | Kepler-56 c | 1.32 | 4.23 | 0.17 | 0.57 |
| HD 180902 b | 1.41 | 4.16 | 1.40 | 1.69 | HIP 75092 b | 1.28 | 4.53 | 2.02 | 1.79 | Kepler-56 d | 1.29 | 4.22 | 2.16 | 5.61 |
| HD 206610 b | 1.55 | 6.12 | 1.74 | 2.04 | HIP 90988 b | 1.30 | 3.94 | 1.26 | 1.96 | Kepler-815 b | 1.25 | 3.42 | 0.09 | 0.05 |
| HD 219415 b | 1.00 | 2.90 | 3.20 | 1.00 | K2-132 b | 1.08 | 3.85 | 0.09 | 0.49 | TOI-2337 b | 1.32 | 3.22 | | 1.60 |
| HD 221416 b | 1.21 | 2.94 | 0.12 | 0.19 | K2-161 b | 0.99 | 2.57 | | 0.10 | TOI-2669 b | 1.19 | 4.10 | | 0.61 |
| HD 222076 b | 1.07 | 4.10 | 1.83 | 1.56 | K2-39 b | 0.66 | 2.97 | 0.06 | 0.09 | | | | | |
| HD 27442 b | 1.23 | 3.18 | 1.27 | 1.56 | K2-97 b | 1.20 | 4.47 | 0.09 | 0.48 | | | | | |